\documentclass[aip, amsmath,amssymb, reprint]{revtex4-1}
\usepackage{dcolumn}
\usepackage{color}
\usepackage{graphicx}
\usepackage{bm}
\usepackage{soul}
\usepackage[utf8]{inputenc}
\usepackage[T1]{fontenc}
\usepackage{mathptmx}
\usepackage{natbib}
\usepackage{multirow}
\usepackage{colortbl}
\usepackage{enumitem}

\definecolor{Grayy}{rgb}{0.9,0.9,0.9}
\definecolor{Mauve}{rgb}{1,0.9,1}
\definecolor{ligthgreen}{rgb}{0.9,1,0.4}

\definecolor{OliveGreen}{rgb}{0,0.6,0}
\definecolor{NavyBlue}{rgb}{0.0, 0.07, 0.8}
\definecolor{ModeBeige}{rgb}{0.59, 0.44, 0.09}
\definecolor{MordantRed}{rgb}{0.8, 0.03, 0.0}
\definecolor{Orange}{rgb}{1.0, 0.5, 0.0}
\definecolor{SkyMagenta}{rgb}{0.81, 0.44, 0.69}
\definecolor{BlueGreen}{rgb}{0.0, 0.7, 0.7}
\definecolor{BlueViolet}{rgb}{0.54, 0.17, 0.89}

\newcommand{\corr}[1]{#1}

\usepackage{amsmath}
\usepackage{amsfonts}
\usepackage{amssymb}

\newcommand{\titre}{Geometrical properties of 3D crossed nanowire networks}

\begin{document}

\title{\titre}
\author{Tristan da C\^{a}mara Santa Clara Gomes}
\affiliation{Institute of Condensed Matter and Nanosciences, Universit\'{e} catholique de Louvain, Place Croix du Sud 1, 1348 Louvain-la-Neuve, Belgium}

\author{Nicolas Marchal}
\affiliation{Institute of Condensed Matter and Nanosciences, Universit\'{e} catholique de Louvain, Place Croix du Sud 1, 1348 Louvain-la-Neuve, Belgium}

\author{Anatole~Moureaux}
\affiliation{Institute of Condensed Matter and Nanosciences, Universit\'{e} catholique de Louvain, Place Croix du Sud 1, 1348 Louvain-la-Neuve, Belgium}

\author{Simon de Wergifosse}
\affiliation{Institute of Condensed Matter and Nanosciences, Universit\'{e} catholique de Louvain, Place Croix du Sud 1, 1348 Louvain-la-Neuve, Belgium}

\author{Chloé~Chopin}
\affiliation{Institute of Condensed Matter and Nanosciences, Universit\'{e} catholique de Louvain, Place Croix du Sud 1, 1348 Louvain-la-Neuve, Belgium}

\author{Luc Piraux}
\affiliation{Institute of Condensed Matter and Nanosciences, Universit\'{e} catholique de Louvain, Place Croix du Sud 1, 1348 Louvain-la-Neuve, Belgium}

\author{Joaqu\'in~de~la~Torre~Medina}
\affiliation{Instituto de Investigaciones en Materiales-Unidad Morelia, Universidad Nacional Autónoma de México, Morelia 58000, Mexico}

\author{Flavio Abreu Araujo}
\email{flavio.abreuaraujo@uclouvain.be}
\affiliation{Institute of Condensed Matter and Nanosciences, Universit\'{e} catholique de Louvain, Place Croix du Sud 1, 1348 Louvain-la-Neuve, Belgium}

\begin{abstract}
Three-dimensional interconnected nanowire networks have recently attracted notable attention for the fabrication of new devices for energy harvesting/storage, sensing, catalysis, magnetic and spintronic applications and for the design of new hardware neuromorphic computing architectures. However, the complex branching of these nanowire networks makes it challenging to investigate these 3D nanostructured systems theoretically. Here, we present a theoretical description and simulations of the geometric properties of these 3D interconnected nanowire networks with selected characteristics. Our analysis reveals that the nanowire segment length between two crossing zones follows an exponential distribution. This suggests that shorter nanowire segments have a more pronounced influence on the nanowire network properties compared to their longer counterparts. Moreover, our observations reveal a homogeneous distribution in the smallest distance between the cores of two crossing nanowires. The results are highly reproducible and unaffected by changes in the nanowire network characteristics. \corr{The density of crossing zones and interconnected nanowire segments are found to vary as the square of the nanowire density multiplied by their diameter, further multiplied by a factor dependent on the packing factor.} Finally, densities of interconnected segments up to 10$^{13}$ cm$^{-2}$ can be achieved for 22-$\mu$m-thick nanowire networks with high packing factors. This has important implications for neuromorphic computing applications, suggesting that the realization of 10$^{14}$ interconnections, which corresponds to the approximate number of synaptic connections in the human brain, is achievable with a nanowire network of about 10 cm$^{2}$.
\end{abstract}

\maketitle

\section{Introduction}

Three-dimensional (3D) networks made of interconnected nanowires (NWs) and nanotubes (NTs) have been developed in the past decades, raising interest due to their unique architecture with high degree of NW interconnectivity, mechanical and functional properties \cite{Rauber2011, Martin2014, Araujo2015, Piraux2020, Camara-Santa-Clara-Gomes2021}. Notably, the system has potential applications in a wide range of fields such as in energy harvesting/storage systems \cite{Wang2012, Wei2013, Antohe2016}, electronic sensing devices and actuators \cite{Kwon2012, Piraux2016, Paulowicz2015}, catalysts \cite{Rauber2011}, electrochromic elements \cite{Scherer2013}, solar cells \cite{Crossland2009}, biosensors \cite{Wang2015}, bio-analytical devices \cite{Rahong2014, Rahong2015}, photonic devices \cite{Martin2014}, \corr{thermal regulators \cite{Liu2024},} magnetic memory \cite{Burks2021}, spintronics\cite{Camara-Santa-Clara-Gomes2016_JAP, Camara-Santa-Clara-Gomes2016_Nanoscale, Pacheco2017, Camara-Santa-Clara-Gomes2021}, thermoelectric devices \cite{Camara-Santa-Clara-GomesAPL, Wagner2021, JPD2022, Nico2023, Luc2023, nano2023}, spin caloritronics devices \cite{Camara-Santa-Clara-Gomes2019, Abreu-Araujo2019, Marchal2020, Tristan2020, Camara-Santa-Clara-Gomes2021_new} and unconventional computing \cite{Bhattacharya2022, Milano2022, Chopin2023}. The NW branching structure offers a good mechanical stability to the networks, that have been found to be self-supported \cite{Camara-Santa-Clara-Gomes2016_JAP, Camara-Santa-Clara-Gomes2016_Nanoscale, Piraux2020}. Moreover, the NW interconnections allow to easily measure electrical and thermoelectrical properties of these nanostructures. However, despite the interest generated by the architecture, theoretical and simulation studies are still lacking to provide a better understanding of 3D crossed NW networks. This is due to the complexity of the nanostructure. Here, we propose theoretical and simulation studies of the geometrical properties of 3D interconnected NW networks with various selected characteristics. 

\begin{figure}[ht!]
    \centering
    \includegraphics[scale=0.6]{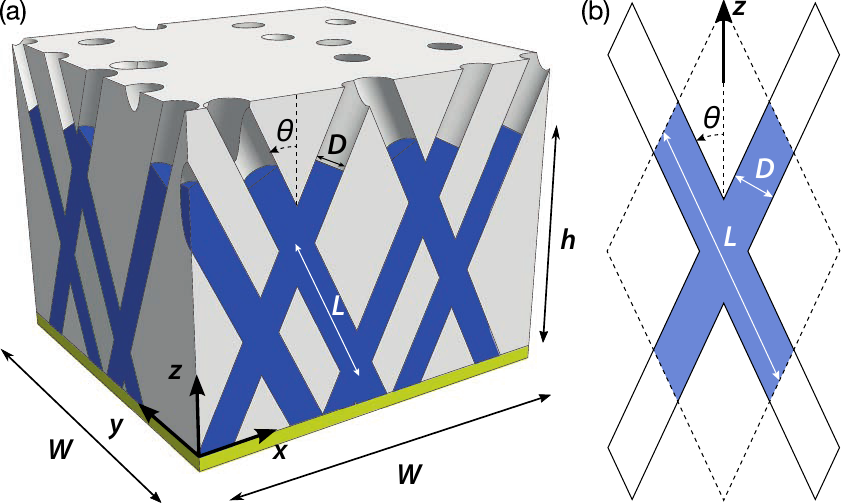}
    \caption{(a) Schematic representation of the simulated volume of 3D nanowire network and its dimensions, $W$ being the side of the network film surface considered and $h$ the network film thickness. The nanowires of diameter $D$ make an angle $\theta$ with the surface normal of the network film, which is the $z$ direction. The $x$ and $y$ directions refer to two perpendicular directions in the plane of the network film surface. The inter-distance between centers of nanowire crossing zones is noted $L$. (b) Unit cell of two crossed nanowires for an inter-distance $L$ between centers of nanowire crossing zones.}
    \label{Fig1}
\end{figure}

3D NW networks are a unique nanoarchitecture made of interconnected NWs to form a mechanically stable and self-supported film. Electrodeposition into well-designed track-etched polymer templates has been proved to be a suitable and low-cost synthesis technique to fabricate 3D nanowire networks with very different geometrical characteristics \cite{Rauber2011, Araujo2015}, such as mean NW diameter $D$, the angle with the network film normal $\theta$, the packing factor $P$ and the film thickness $h$. Here, we propose the study of NW networks film with characteristics corresponding to recent studies by Piraux et al. \cite{Piraux2020}. In this system, NWs are grown into well-defined templates with crossed cylindrical nanochannels, which are created by exposing a polycarbonate film to four irradiation steps with different incident angles. Considering the $z$ direction as the normal direction of the film, and the $x$ and $y$ directions as two orthogonal in-plane directions of the film, as illustrated in Figure \ref{Fig1}(a), the four resulting nanopore families make angles of $-\theta$ and $+\theta$ with respect to the $z$-direction in the plane $x$-$z$ and $y$-$z$, with $\theta$ encompassed between 20$^\circ$ and 25$^\circ$. We will consider different high porosity systems with diameters of $D \approx$ 20 nm, 40 nm, 105 nm and 230 nm and respective ion impact surface densities $n$ of 4.8$\times$10$^{10}$ cm$^{-2}$, 1.2$\times$10$^{10}$ cm$^{-2}$, 2.4$\times$10$^{9}$ cm$^{-2}$ and 5.0$\times$10$^{8}$ cm$^{-2}$. The porosity of the template can be estimated as $P = n \pi D^2/4$, which yields respectively $P \approx$ 15\%, 15\%, 21\% and 21\% for the high porosity systems studied. Additionally, we will consider two low porosity systems with $D \approx$ 40 nm and 80 nm, and $n =$ 6$\times$10$^{8}$ cm$^{-2}$, giving respectively $P \approx$ 0.8\% and 3\%. \corr{Note that the porosity of the template is assumed to be identical to the NW network packing factor, indicating that the entire template is considered as filled with NWs.}

\begin{figure*}[ht!]
    \centering
    \includegraphics[scale=0.7]{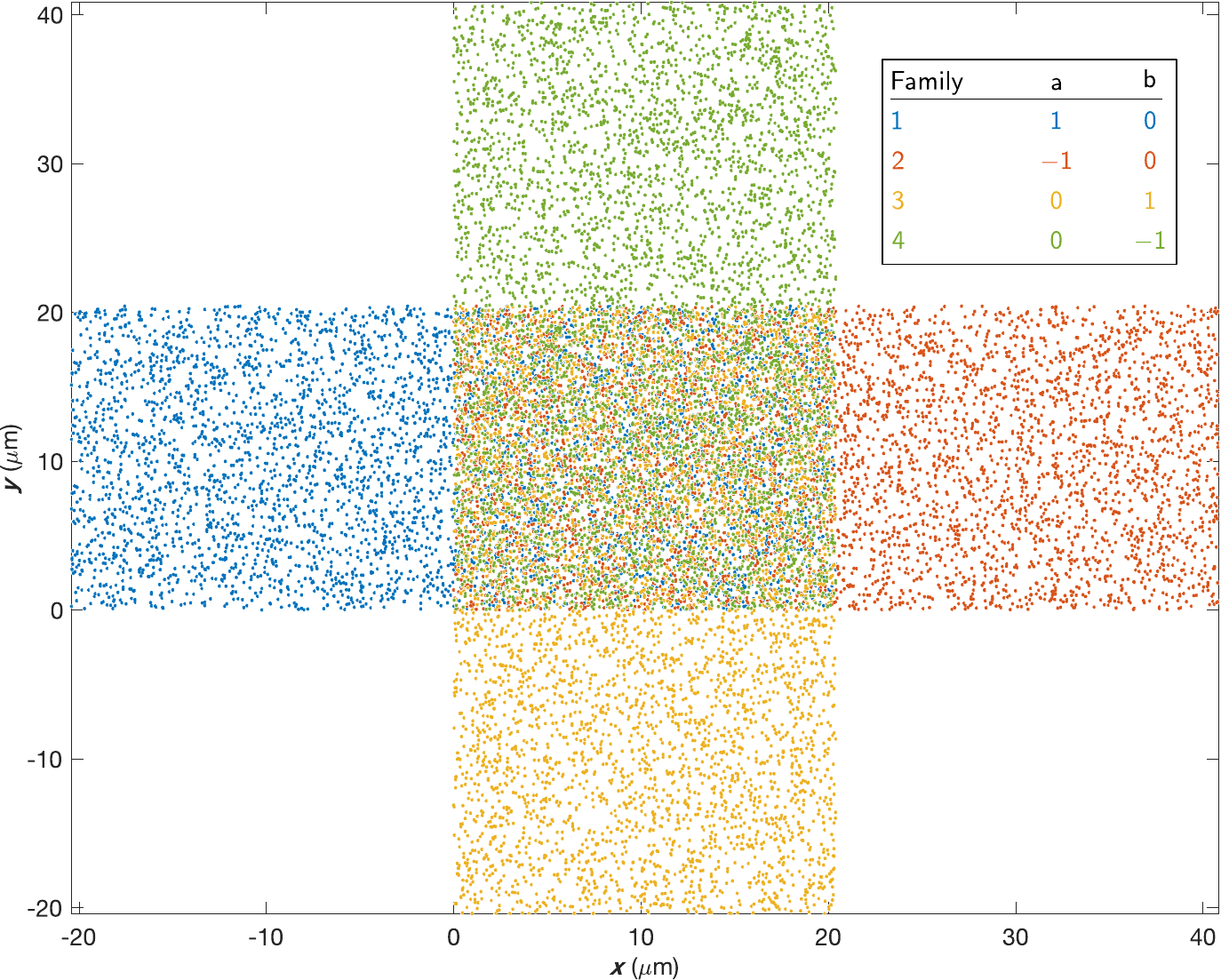}
    \caption{Schematic representation of the nanowire network film surface ($z =$ 0) along with the irradiation impacts randomly generated for the characterization of the geometry of the interconnected structure. Each dot represents one impact, the 4 families corresponding to the 4 different colours. The inset provides the values of the constants $a$ and $b$ needed for the orientation vectors for each of the four nanowire families.}
    \label{surface}
\end{figure*}

The complex nanoarchitecture is simulated in a specified volume $V = W^2 h$ of the interconnected NW structure, where $W$ is the side of a square on the plan ($x$, $y$) of the film surface and $h$ is the film thickness along the $z$ direction. This specified volume directly depends on the number of surface irradiation impacts $N = 4 N_f$ selected, where $N_f$ is the number of surface impacts for each of the four families of nanopores. These four families are defined as follows:
\begin{itemize}[noitemsep,topsep=0pt]
    \item Family 1: making an angle of $+\theta$ with the $z$-direction in the plane $x$-$z$;
    \item Family 2: making an angle of $-\theta$ with the $z$-direction in the plane $x$-$z$;
    \item Family 3: making an angle of $+\theta$ with the $z$-direction in the plane $y$-$z$;
    \item Family 4: making an angle of $-\theta$ with the $z$-direction in the plane $y$-$z$.
\end{itemize}
The studied volume is illustrated in Figure \ref{Fig1}(a) with the different parameters. Besides, Figure \ref{Fig1}(b) shows the NW crossing unit cell, which depends on the length $L$ between the centers of two successive NW interconnections. Using the surface density of impact $n$, the length $W$ is obtained as $W = \sqrt{N/n}$, while the thickness is selected by assuming that if a NW begins at one edge of the surface, its end is located on the opposite edge, which gives $h = W/\tan{\theta}$. Taking the thickness $h$ as fixed by the system, it yields a number of surface irradiation impacts $N = n (h\tan{\theta})^2$, which is rounded to the closest multiple of four to account for the four identical irradiation for each of the four families. The center of the irradiation impact ($x_0$, $y_0$) of each nanopore from the four families on the surface $z=$ 0 is randomly chosen. Figure \ref{surface} shows the bottom surface ($z =$ 0) for $D =$ 40 nm, $n =$ 1.2$\times$10$^{10}$ cm$^{-2}$ and $h =$ 22 $\mu$m, where each of the four nanopore families are indicated in different colours. The NW segments are simulated by cylinders of diameter $D$ along straight lines \corr{(referred as NW axes)} with orientation vector $(a \sin \theta, b \sin \theta, \cos \theta)$, where $a$ and $b$ are constants that depend on their family, as provided in the inset of Figure \ref{surface}. Moreover, NWs engendered by impacts from outside the considered surface can also induce crossings inside the considered volume. Therefore, NWs from each family are also created from $N$ additional impacts randomly chosen outside of the surface as shown in Figure \ref{surface}. The NWs originating from inside the surface of the volume of interest are noted as "in", while the NWs originating from outside the surface of the volume of interest are noted as "out". This virtual NW network has been used for all the following characterizations.

\section{Nanowire segment length distribution}


\corr{In this section, we present the derivation of the statistical distribution of the inter-distance between two successive crossing points, noted $L$, obtained through simulations. A crossing point is defined as the center of the interconnection volume between two NWs. Understanding the distribution of $L$ is essential for simulating the structure of crossed NWs, as it plays a crucial role in defining the unit cell of the crossed NW configuration, which is illustrated in Figure \ref{Fig1}b. In addition, we derive the shortest distance between the axis of two NWs, denoted as $d$. This distance varies from zero for NWs with axis intersecting, to $\pm D/2$ for NWs that only touch each other. Furthermore, we investigate the densities of interconnections and NW segments, as well as the proportion of interconnections involving two, three, or more NWs.} For this approach, the crossing conditions are separated into three main categories: 
\begin{itemize}[noitemsep,topsep=0pt]
    \item Crossing between two NWs generated from impacts at the bottom surface of the considered volume
    \item Crossing between a NW generated from impacts at the bottom surface of the considered volume and a NW generated from impacts outside of the bottom surface of the considered volume
    \item Crossing between two NWs generated from impacts outside of the bottom surface of the considered volume
\end{itemize}
\corr{Note that we exclude the possibility of interconnection between two nanowires belonging to the same family, as the angle is considered as fixed. Therefore, the length between two crossing points is expected to be slightly overestimated.}

\corr{We analyze each pair of simulated NW axes, represented by orientation vectors $(a_1 \sin \theta, b_1 \sin \theta, \cos \theta)$ and $(a_2 \sin \theta, b_2 \sin \theta, \cos \theta)$ for the first and second NWs considered, respectively, along with their respective irradiation impact centers ($x_0$, $y_0$)$_1$ and ($x_0$, $y_0$)$_2$. We always designate the first NW as the one with the smaller family index. The shortest distance $|d|$ between two NWs' axis is computed using the formula: }
\begin{equation}
    d = \left|\frac{(b_1-b_2)\Delta x - (a_1-a_2)\Delta y}{\sqrt{K}}\right|
    \label{EqD}
\end{equation}
where $K = (a_1 - a_2)^2 + (b_1 - b_2)^2 + (a_1^2b_2^2 + a_2^2b_1^2)\tan^2\theta$ \corr{and the difference between the center of impacts ($\Delta x$, $\Delta y$) is defined as ($x_0$, $y_0$)$_1$ $-$ ($x_0$, $y_0$)$_2$. A crossing point is considered only if this distance $|d|$ is smaller that the diameter $D$ and is located within the total volume ($W\times W\times h$) considered in our simulations. The computed values of $K$ and $|d|$ for each pair of families are reported in Table S1 in the Supplementary Material \cite{SuppMat} Section S1, together with the detailed conditions for crossing (see Table S2). }

The position of the crossing \corr{point} on the two NWs \corr{axes}, which are noted $P_1$ and $P_2$\corr{,} are computed as
\begin{equation}
    P_1 = \frac{-\Delta x (a_1 - a_2 + b_2K_1) - \Delta y (b_1 - b_2 + a_2K_2)}{K\sin\theta}
    \label{EqsP}
\end{equation}
and
\begin{equation}
    P_2 = \frac{-\Delta x (a_1 - a_2 - b_1K_2) - \Delta y (b_1 - b_2 - a_1K_1)}{K\sin\theta}\text{,}
    \label{EqsP2}
\end{equation}
with $K_1 = a_1b_2\tan^2\theta$ and $K_2 = a_2b_1\tan^2\theta$. \corr{After computing the crossing positions $P_i$ for each NW, the inter-distances $L$ between two crossing \corr{points} are obtained by computing the distances between successive crossing positions $P_i$.} 

\begin{figure*}[ht!]
\centering
\includegraphics[width = 0.9 \textwidth]{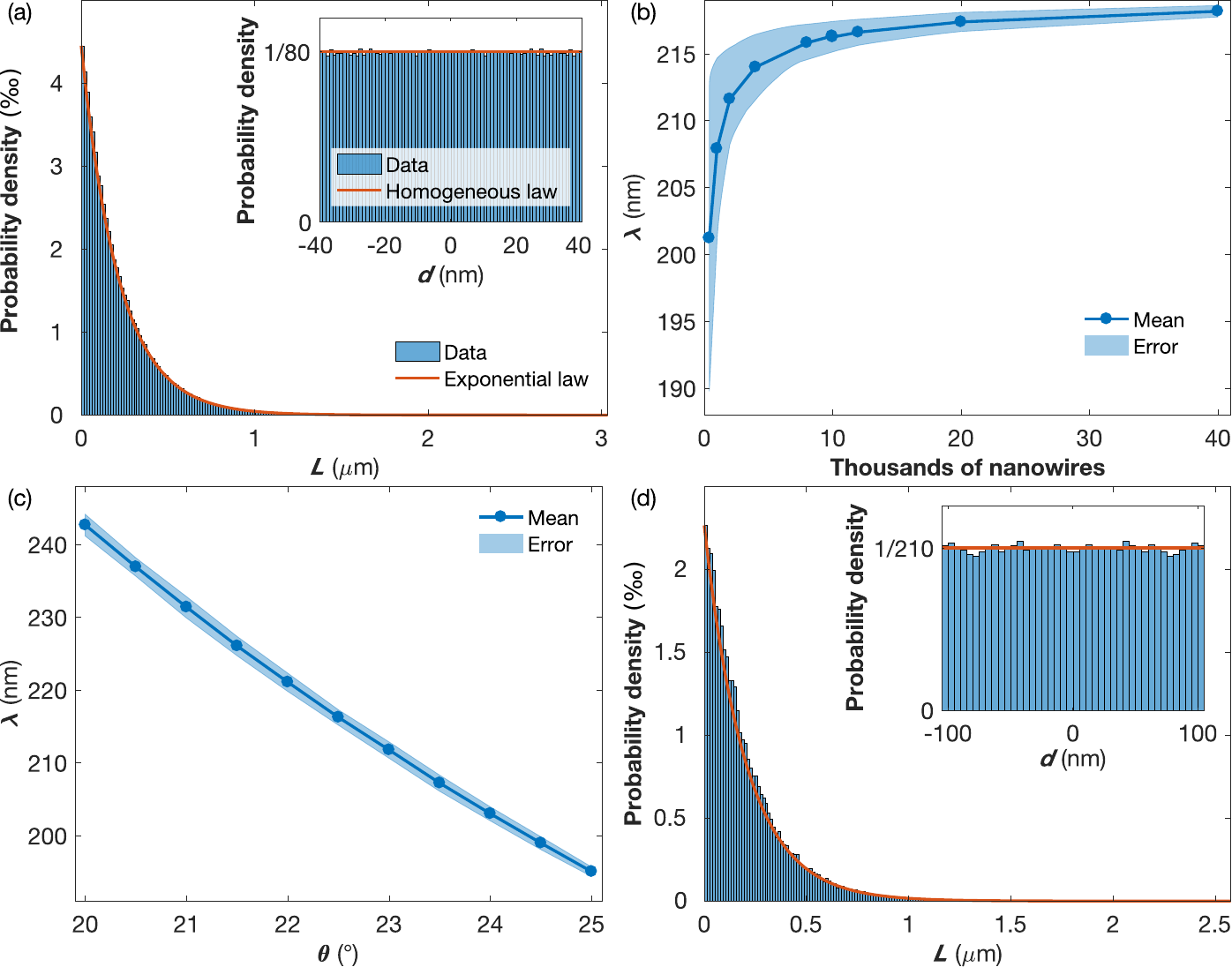}
\caption{(a) Probability density for the length between two successive crossing \corr{points}, showing the simulated data (in blue) for $D =$ 40 nm, $\theta$ = 22.5$^\circ$, $n =$ 1.2$\times$10$^{10}$ cm$^{-2}$ and $h =$ 22 $\mu$m, and an exponential law $1/\lambda \exp(l/\lambda)$ (in red) with $\lambda$ taken as the mean length between two successive crossing \corr{points}. The inset shows the corresponding probability density for the shortest distance between two crossing nanowires \corr{axes} with the same parameter, including the simulated data (in blue) and a homogeneous law $1/(2D)$. (b) Mean length between two successive crossing \corr{points} as a function of the number of simulated nanowires (in thousands) taken as the mean value over one hundred simulations for a nanowire network with 40 nm in diameter and about 15\% in \corr{packing factor}. (c) Mean length between two successive crossing \corr{points} as a function of the angle $\theta$ taken as the mean value over one hundred simulations for a nanowire network with 40 nm in diameter and about 15\% in \corr{packing factor}. The blue areas in (b) and (c) show the error taken as 3 times the standard deviation over one hundred runs. (d) Probability density for the length between two successive crossing \corr{points}, showing the simulated data (in blue) for $D =$ 105 nm, $\theta$ = 22.5$^\circ$, $n =$ 2.4$\times$10$^{9}$ cm$^{-2}$ and $h =$ 22 $\mu$m, and an exponential law $1/\lambda \exp(l/\lambda)$ (in red) with $\lambda$ taken as the mean length between two successive crossing \corr{points}. The inset shows the corresponding probability density for the shortest distance between two crossing nanowires \corr{axes} with the same parameters, including the simulated data (in blue) and a homogeneous law $1/(2D)$.}
\label{FigProb}
\end{figure*}

First, a NW network with $D =$ 40 nm, $\theta$ = 22.5$^\circ$, $n =$ 1.2 10$^{10}$ cm$^{-2}$, $W = $ 9.1 $\mu$m, $h =$ 22 $\mu$m, which corresponds to $N =$ 9964 NWs, is simulated. \corr{A set of one thousand simulations of the system generates a mean value of about 538 000 crossing points (with a standard deviation of about 0.2\%)}, which corresponds to about \corr{294 crossing points} per $\mu$m$^3$. The distribution of the inter-distance between two centers of successive NWs interconnections $L$ is shown in Figure \ref{FigProb}(a) for this network. It shows an exponential distribution with a mean value of $\lambda =$ 216 nm (this value is identical to the standard deviation \corr{of the exponential distribution}). \corr{This value fluctuates by approximately 0.3\% across a set of one thousand simulations.} As seen in Figure \ref{FigProb}(a), the theoretical exponential law $1/\lambda \exp(l/\lambda)$ corresponds well to the simulated data. The largest inter-distance between two crossing points is slightly larger than 3 $\mu$m. In addition, we observe that 99\% of the segments between two crossings points are below 1 $\mu$m. 
Besides, we find that 60\% of the crossing \corr{points} involve perpendicular families. In addition, the inset of Figure \ref{FigProb}(a) shows the distribution of the shortest distance $d$ between two crossing NWs' \corr{axes}. As seen, it displays a homogeneous distribution between $-$40 nm and 40 nm, which corresponds to the value of the diameter $D$ as expected as two NWs can only cross if the shortest distance between them is less than the diameter. These simulations are repeated several times with very similar results at each run (the mean value and standard deviation of $\lambda$ for one hundred runs with these parameters are 216.2 nm and 0.4 nm). Figure \ref{FigProb}(b) shows the mean value of $\lambda$ as a function of the number $N$ of simulated irradiation impacts on the surface of the simulated volume while keeping the same surface density of impacts $n$, obtained on one hundred runs for each value of $N$. The blue area shows the error taken as three times the standard deviation for these one hundred runs (gathering 99.7\% of the variation). As seen, the mean value converges when the value of $N$ is increased while the error decreases as expected from the law of large numbers.\\

\begin{figure*}[ht!]
\centering
\includegraphics[width = 0.9 \textwidth]{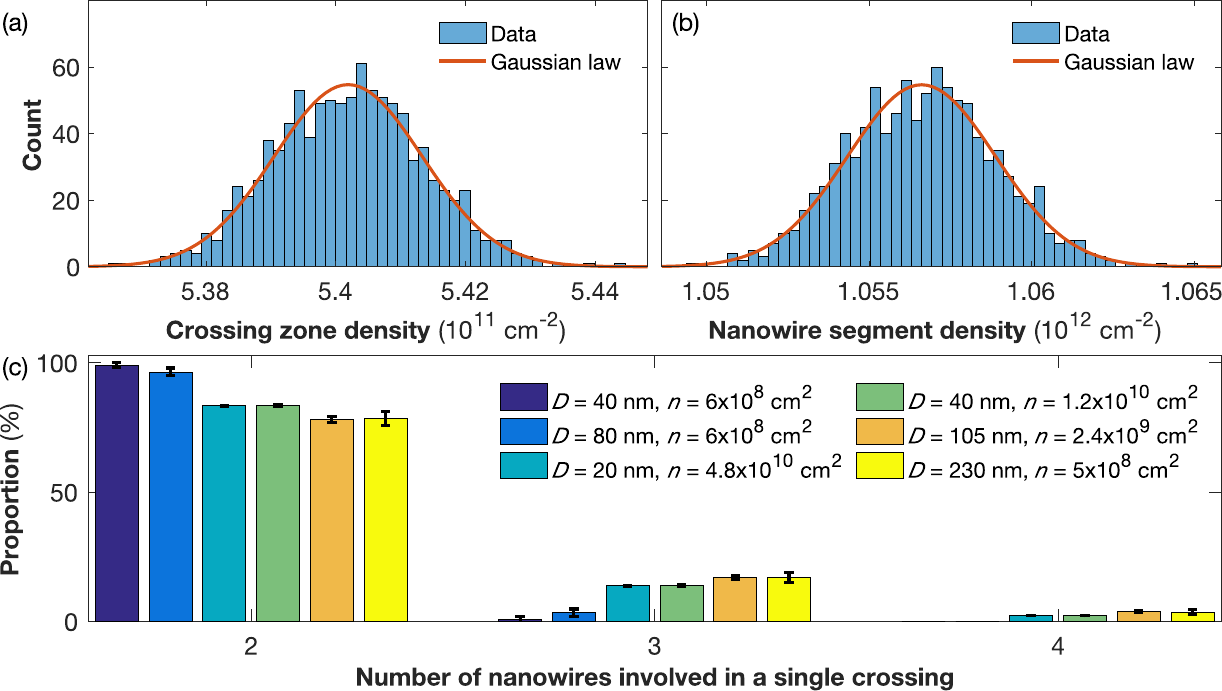}
\caption{\corr{(a-b) Count of the (a) crossing zone and (b) nanowire segment densities obtained from one thousand simulations of nanowire networks with $D =$ 40 nm, $\theta$ = 22.5$^\circ$, $n =$ 1.2$\times$10$^{10}$ cm$^{-2}$ and $h =$ 22 $\mu$m, along with the Gaussian distribution plotted using the mean density and their respective standard deviations (see Table \ref{NumLeg}). (c) Proportion of crossing zones involving two, three and four nanowires for various network characteristics.}}
\label{Fig4}
\end{figure*}

In the real system, the angle $\theta$ varies between 20$^\circ$ and 25$^\circ$, which is expected to modify the mean length between two successive crossing \corr{points}. Indeed, a change in $\theta$ is expected to modify the probability of crossing NWs from other families, as well as to induce additional crossings between NWs of the same family. However, this largely complicates the simulation. Instead of randomly affect an angle $\theta$ between 20$^\circ$ and 25$^\circ$ to each NW, the value of $\theta$ is taken as the same value for all NWs and varied from 20$^\circ$ to 25$^\circ$, while the \corr{packing factor} of the NW networks is maintained by adapting the studied volume dimensions. This gave a first step towards understanding the effect of $\theta$ on the NW networks properties. Figure \ref{FigProb}(c) provides the mean value of $\lambda$ for one hundred runs as a function of $\theta$, while the others parameters are kept unchanged ($D =$ 40 nm and $n =$ 1.2$\times$10$^{10}$ cm$^{-2}$). The blue area in Figure \ref{FigProb}(c) provides the error taken as three times the standard deviation. As expected, the larger the value of $\theta$ the lower the mean distance between two crossing points $\lambda$. Indeed, the larger the angle $\theta$, the larger the probability that NW segments cross inside the volume. This also clearly appeared in the increase of the number of crossing \corr{points} going from 362 789 (standard deviation of 806 on one hundred runs) for $\theta =$ 20$^\circ$ to 77 200 (standard deviation of 1044 on one hundred runs) \corr{for $\theta =$ 25$^\circ$}. For all values of $\theta$ considered, the exponential distribution $1/\lambda \exp(l/\lambda)$ is found to fit very well the data, as in Figure \ref{FigProb}(a) for $\theta =$ 22.5$^\circ$. Moreover, the same homogeneous distribution 1/80 is obtained for all values of $\theta$ for the shortest distance $d$ between two crossing NWs' \corr{axes}. 

Similar studies have also been performed for the NW networks with various diameters and ion impact surface density achievable in laboratories, giving very similar results. As an example, Figure \ref{FigProb}(d) shows the exponential distribution of $L$ obtained for a NW network with $D =$ 105 nm of diameter, a surface density of impact $n =$2.4$\times$10$^{9}$ cm$^{-2}$ and a network film thickness $h =$ 22 $\mu$m, while the angle $\theta$ is taken as 22.5$^\circ$. For these parameters, the impact surface has a side $W$ of 9.1 $\mu$m and $N =$ 1992 impacts. Here again, the exponential law $1/\lambda \exp(l/\lambda)$ corresponds very well to the data, where a mean value of $\lambda$ over one \corr{thousand} runs is 406 nm \corr{$\pm$ 3.6} nm. Besides, it is observed that 99\% of the segments between two crossings points are below 2 $\mu$m. A set of one \corr{thousand} simulations \corr{provides} a mean value of \corr{about 56 000 crossing points} (with standard deviation of \corr{about 0.5\%}) in the volume of interest, which corresponds to about 31 \corr{crossing points} per $\mu$m$^3$. Here again, 60\% of the crossing \corr{points} involve NWs from perpendicular families.
Moreover, as seen in the inset of Figure \ref{FigProb}(d), the shortest distance $d$ between two crossing NWs' \corr{axes} displays a homogeneous probability density of $1/(2D)$ as obtained for $D =$ 105 \corr{nm} in the inset of Figure \ref{FigProb}(a). The exponential probability density for $L$ and the homogeneous probability density for $d$ remain when changing the value of $\theta$, while only the value of $\lambda$ is affected. As for the system with 40 nm in diameter, the larger the value of $\theta$, the smaller the value of $\lambda$. 

\corr{One advantageous property of the 3D NW network system is the dense interconnectivity of the NW segments. To describe this characteristic, we investigate the density of NW segments, noted $n_\text{seg}$, and the density of crossing zones between the NW segments, noted $n_\text{cross}$, for NW networks with various diameters $D$ and impact densities $n$, while fixing the network thickness $h$ to 22 $\mu$m. In contrast to a crossing point that is defined for each pair of NWs that are crossing each other, a crossing zone is defined as a volume where multiple NWs interconnect, while the NW segments as the segments between crossing zones. Therefore, the density of crossing points is adjusted to accommodate crossing zones involving multiple crossing points (i.e., more than two NWs interconnecting; refer to Supplementary Material \cite{SuppMat} Section S2 for details). In Figure \ref{Fig4}(a-b), we show the count histograms for the crossing zone density (a) and NW segment density (b) obtained from one thousand simulations of the system with $D =$ 40 nm, $\theta =$ 22.5$^\circ$, $n =$ 1.2$\times$10$^{10}$ cm$^{-2}$ and $h =$ 22 $\mu$m. The results are well fitted by Gaussian distributions, plotted using the calculated mean and standard deviation values from the simulations. For the case considered, we obtain $n_\text{cross} =$ 5.4$\times$10$^{11}$ ($\pm$0.2\%) and $n_\text{seg} =$ 1.1$\times$10$^{12}$ ($\pm$0.2\%). Note that this small standard deviation is anticipated to significantly increase for minor variations in the parameters $h$, $\theta$, $n$ or $D$ that can be expected in real NW network.} Table \ref{NumLeg} provides \corr{the values for $n_\text{cross}$ and $n_\text{seg}$}, as well as the mean distance between two crossing points $\lambda$\corr{,} for the different NW networks simulated with $h$ fixed to 22 $\mu$m. \corr{Similar Gaussian distributions are obtained for all the studied NW networks (see Supplementary Material \cite{SuppMat} Section S2 for the distributions, as well as the mean and standard deviation values for $n_\text{cross}$ and $n_\text{seg}$). Our simulations with various NW network parameters $n$ and $D$ reveals that the density of crossing points between pairs of NWs varies as $n^2D$ (see Supplementary Material \cite{SuppMat} Section S2). This relationship can be attributed to the fact that the number of NWs per cm$^2$ is directly proportional to $n$. Assuming that the probability of two NWs to cross is independent of the number of NWs, then the number of crossing points made by each of the $n$ NWs will be proportional to the number $n$ of NWs, i.e., the density of crossing points is proportional to $n^2$. Additionally, the probability of crossing points is found to be proportional to $D$, as suggested by the condition for crossing $|d| \leq D$. Then, the adjustment made to calculate the density of crossing zones results in a decrease compared to the density of crossing points. This corrective factor is found to be proportional to the network packing factor $P$ (i.e., proportional to $nD^2$), as an increasing packing factor induces an increasing proportion of crossing zones involving more than two NWs (see Supplementary Material \cite{SuppMat} Section S3). Importantly}, NW networks \corr{with $D =$ 20 nm and $n =$ 4.8$\times$10$^{10}$ cm$^{-2}$} exhibit up to about 10$^{13}$ interconnected segments per cm$^{-2}$. This has high interest for neuromorphic computing applications. Indeed, the human brain is estimated to host in the range of 10$^{14}$ synaptic connections \cite{shepherd2003synaptic, koch2004biophysics, Drachman2005, Saver2006, Azevedo2009, Motta2019}. Interestingly, 10$^{14}$ interconnections can be achieved with a network of about 10 cm$^{2}$. \corr{Note that in the real system, the angle $\theta$ varies between 20$^\circ$ and 25$^\circ$, thereby allowing for additional crossings between NWs from the same families. Consequently, the values reported in Table I are expected to be underestimated.}

\begin{table}[h]
   \caption{\label{NumLeg} \corr{Mean crossing zone and nanowire segments densities, $n_\text{seg}$ and $n_\text{cross}$,} and mean distance between two crossing points $\lambda$\corr{, obtained over one thousand simulations of nanowire} network films with $h =$ 22 $\mu$m and various combination of NW diameter $D$ and ion impact surface density $n$.}
\begin{tabular}{ccccc}
  \hline
$D$ (nm) & $n$ (cm$^{-2}$) & $n_\text{cross}$ (cm$^{-2}$) & $n_\text{seg}$ (cm$^{-2}$) & $\lambda$ (nm)\\
    \hline
20 & 4.8$\times$10$^{10}$ & \corr{4.3}$\times$10$^{12}$ & \corr{8.6$\times$10$^{12}$} & 109 \\
40 & 1.2$\times$10$^{10}$ & \corr{5.4}$\times$10$^{11}$ & \corr{1.1}$\times$10$^{12}$ & 216\\
105 & 2.4$\times$10$^{9}$ & \corr{5.3}$\times$10$^{10}$ & \corr{1.0}$\times$10$^{11}$ & 405 \\
230 & 5$\times$10$^{8}$ & \corr{5.0}$\times$10$^{9}$ & \corr{9.0}$\times$10$^{9}$ & 883\\
40 & 6$\times$10$^{8}$ & \corr{1.6}$\times$10$^{9}$ & \corr{2.2}$\times$10$^{9}$ & 2740\\
80 & 6$\times$10$^{8}$ & \corr{3.1}$\times$10$^{9}$ & \corr{5.2}$\times$10$^{9}$ & 1770 \\
  \hline
\end{tabular}
\end{table}

\corr{In Figure \ref{Fig4}(c), we present the proportion of crossing zones involving two, three or four NWs for various networks with given $n$ or $D$ characteristics. As depicted, the proportion of crossing zones involving more than two NWs increases with increasing $n$ or $D$. Notably, we observe that the proportion of crossing zone involving two NWs is directly dependent on the NW network packing factor (See Supplementary Material \cite{SuppMat} Section S3 for further details). Note that crossing zones can involve more than four NWs in the systems with $D =$ 105 nm and $n =$ 2.4$\times$10$^{9}$ cm$^{-2}$ and $D =$ 230 nm and $n =$ 5$\times$10$^{8}$ cm$^{-2}$. However, these crossing zones account for less than 1\% of the total crossing zones in the system.}

\corr{These preliminary results can be used as a base to study physical properties of interest within complex 3D network of NWs with dense interconnectivity, such as electrical or heat transport, magnetism, or temperature distribution. This by simulating various smaller and simpler NW crossing configurations of the system and utilizing the statistical properties derived here to reconstruct the anticipated behavior of the entire system. It should be noticed that 3D NW network systems remain very complex nanostructures and additional imperfections such as defect in the template or inhomogeneous growth can impact the structure.}

\section{conclusion}

In this study, we characterise analytically some geometric properties of three-dimensional interconnected nanowire networks with selected characteristics. We observe an exponential distribution of the nanowire segment size between two interconnections, indicating a very large number of short nanowires. Our investigation reveals that altering the angle $\theta$ affects the network's properties. As $\theta$ increased, the mean inter-distance between two crossing \corr{points} decreases, aligning with the increased probability of NW segments crossing within the volume. This trend is consistent across varying values of $\theta$, with the exponential distribution fitting the data effectively. Simulations of networks with different nanowire diameters and ion impact surface densities also provide similar results. \corr{The densities of crossing zones and interconnected nanowire segments are proportional to $n^2 D$ multiplied by a factor dependent on the nanowire packing factor $P = n \pi D^2/4$, where $n$ and $D$ refer to the ion impact surface density and nanowire diameter, respectively.} Importantly, these NW networks demonstrate the potential for achieving up to about 10$^{13}$ interconnected segments per cm$^{2}$, a crucial factor for applications in neuromorphic computing. Notably, our findings suggest that a network covering approximately 10 cm$^{2}$ could achieve the significant 10$^{14}$ synaptic connections estimated in the human brain. To conclude, our analytical study provides valuable insights into the geometric properties and interconnections within three-dimensional nanowire networks. These findings contribute to our understanding of the impact of key parameters on network characteristics and highlight the potential of such networks for applications requiring dense interconnectivity.\\

\noindent
\textbf{Acknowledgments}\\
Financial support was provided by the Belgian Fund for Scientific Research (F.R.S.-FNRS) and F.A.A. is a Research Associate of the F.R.S.-FNRS. S.d.W. is a FRIA grantee, also from the F.R.S.-FNRS. J.d.l.T.M. thanks CONAHCYT for financial support through project A1-S-9588. \\

\noindent
\textbf{Authors contribution}\\
F.A.A. supervised, conceived and design the study. F.A.A. and T.d.C.S.C.G. developed the model code and performed the simulations. T.d.C.S.C.G., F.A.A., and N.M. analysed the data. T.d.C.S.C.G. prepared the manuscript. All authors discussed the results and contributed to the final manuscript.\\

\section*{Reference}

\bibliographystyle{unsrt}



\end{document}